\documentclass[english,reqno]{smfart}
\usepackage{epsfig,times}
\usepackage[mathcal]{euscript}
\usepackage{amsmath}
\usepackage{amsthm}
\usepackage{amsfonts}
\usepackage{amssymb}
\usepackage{masse}
\newcommand{\fr}[2]{{\textstyle \frac{#1}{#2} }}
\newcommand{\ga}{\gamma}
\newcommand{\CA}{{\mathcal A}}

\newcommand{\CH}{{\mathcal H}}

\newcommand{\CU}{{\mathcal U}}
\newcommand{\CV}{{\mathcal V}}

\newcommand{\SL}{{\mathsf L}}

\newcommand{\fsl}{{\mathfrak s}{\mathfrak l}}

\newcommand{\sfc}{{\mathsf c}}

\renewcommand{\sf}{{\mathsf f}}

\newcommand{\sll}{{\mathsf l}}

\newcommand{\sq}{{\mathsf q}}
\newcommand{\spp}{{\mathsf p}}

\newcommand{\sx}{{\mathsf x}}

\newcommand{\sz}{{\mathsf z}}

\newcommand{\BR}{{\mathbb R}}

\newcommand{\BS}{{\mathbb S}}

\def\beq{\begin{equation}}                     %
\def\eeq{\end{equation}}                       %
\def\bea{\begin{eqnarray}}                     
\def\eea{\end{eqnarray}}                       
\def\NP{{\it Nucl. Phys.} }                    
\def\PL{{\it Phys. Lett.} }                    
\begin {document}                 
\title{Quantum Liouville theory versus quantized
Teichm\"uller spaces}
%
%
%
%
%
%
\author{
%
%
%
%
%
J. Teschner                              
}
\address{
%
%
%
%
Institut f'\"ur theoretische Physik\\       %
Freie Universit\"at Berlin,\\               
Arnimallee 14\\                             
14195 Berlin\\ Germany                      
}
\begin{abstract}
%
%
This note announces the proof of a conjecture of
H. Verlinde, according to which the spaces of 
Liouville conformal blocks and the Hilbert spaces from the 
quantization of the
Teichm\"uller spaces of Riemann surfaces carry equivalent
representations of the mapping class group. 
This provides a basis for the geometrical interpretation
of quantum Liouville theory in its relation 
to quantized spaces of Riemann surfaces.   

\vspace*{2mm}\noindent
Contribution to the proceedings of the 35th Ahrenshoop Symposium, 2002
\end{abstract}
\maketitle
\section{Introduction}

Quantum Liouville theory is a crucial ingredient for a variety 
of models for low dimensional quantum gravity
and noncritical string theories. In the case of two dimensional 
quantum gravity or noncritical string theories this comes about 
due to the Weyl-anomaly \cite{Pol}, which forces one to 
take into account the quantum dynamics of the conformal factor of the
two-dimensional metric. More recently it was proposed that 
Liouville theory also plays a crucial role for {\it three-dimensional} quantum
gravity in the presence of a cosmological constant in the sense of
representing a holographic dual for this theory, see e.g.
\cite{Kr,KV}.

For all these applications it is crucial that the quantum Liouville theory
has a geometric interpretation as describing the  quantization of spaces
of two-dimensional metrics. Such an interpretation is to be expected
due to the close connections between {\it classical} Liouville theory and 
the theory of Riemann surfaces. Having fixed a complex structure on the
Riemann surface one may represent the 
unique metric of negative constant curvature locally in the form 
$ds^2=e^{2\varphi}dzd\bar{z}$ where $\varphi$ must solve the
Liouville equation $\partial\bar{\partial}\varphi=\frac{1}{4}e^{2\varphi}$.
The relation between the Liouville equation and the uniformization 
problem leads to beautiful connections between classical Liouville theory
and the theory of moduli spaces of Riemann surfaces \cite{ZT}.

There has been considerable recent progress on the Liouville quantum field
theory as a conformal field theory in its own right, see \cite{TL} 
and references therein. What was missing so far is the
interpretation of these results as a description of the 
quantum geometry of Riemann surfaces. The results that we want to 
present here may be seen as representing the ``chiral half'' of
such an interpretation. More precisely, we would like to announce the proof
of a conjecture of H. Verlinde \cite{V}, according to which the space of 
conformal blocks of the Liouville theory with its mapping class 
group representation is isomorphic to the 
space of states obtained by quantizing the Teichm\"uller spaces of
Riemann surfaces \cite{Fo,Ka1,CF}.

\section{Teichm\"uller spaces} 

The Teichm\"uller spaces are the spaces
of deformations of the complex structures on Riemann surfaces.
As there is a unique metric of constant curvature -1 associated to 
each complex structure one may identify the Teichm\"uller spaces
with the spaces of deformations of the metrics with constant curvature -1.
Coordinates for the Teichm\"uller spaces can therefore be obtained
by considering the geodesics defined by the constant curvature metrics.

\begin{figure}
\begin{center}\epsfxsize7cm
\epsfbox{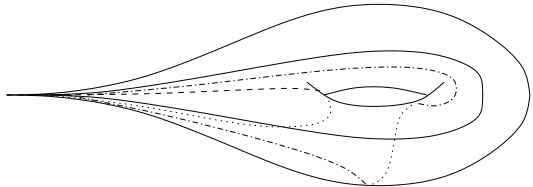}\hspace{1cm}
\end{center}
\caption{Triangulation of the once-punctured torus.}\label{triang}
\end{figure}

A particularly useful set of coordinates was introduced
by R. Penner in \cite{Pe}. They can be defined for Riemann surfaces 
that have at
least one puncture. One may assume having triangulated the surface 
by geodesics that start and end at the punctures. As an example we have 
drawn in Figure \ref{triang} a triangulation of
the once-punctured torus.
The length of these
geodesics will be infinite. In order to regularize this divergence
one may introduce one horocycle around each puncture and measure 
only the length of the segment of a geodesic that lies between the 
horocycles. Assigning to an edge $e$ its regularized length $l_e$
gives coordinates for the so-called decorated Teichm\"uller spaces. 
These are fibre spaces over the
Teichm\"uller spaces which have fibres that parametrize the choices of the 
``cut-offs'' as introduced by the horocycles. 

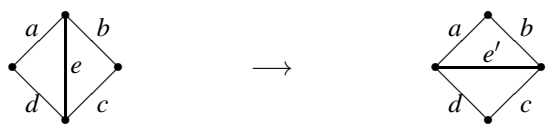
\begin{figure}[h]
\centering
\begin{picture}(200,40)
\put(20,0){\line(-1,1){20}}\put(5,3){$d$}
\put(20,0){\line(1,1){20}}\put(32,3){$c$}
\put(0,20){\line(1,1){20}}\put(5,32){$a$}
\put(40,20){\line(-1,1){20}}\put(32,32){$b$}
\put(20,0){\line(0,1){40}}\put(22,18){$e$}
\put(180,0){\line(-1,1){20}}\put(165,3){$d$}
\put(180,0){\line(1,1){20}}\put(192,3){$c$}
\put(160,20){\line(1,1){20}}\put(165,32){$a$}
\put(200,20){\line(-1,1){20}}\put(192,32){$b$}
\put(160,20){\line(1,0){40}}\put(178,22){$e'$}
\put(20,0){\circle*{3}}
\put(0,20){\circle*{3}}
\put(20,40){\circle*{3}}
\put(40,20){\circle*{3}}
\put(180,0){\circle*{3}}
\put(160,20){\circle*{3}}
\put(180,40){\circle*{3}}
\put(200,20){\circle*{3}}
\put(90,17){$\longrightarrow$}
\end{picture}
\caption{The elementary move between two triangulations}\label{fliplab}
\end{figure}

A closely related set of coordinates for 
the Teichm\"uller spaces themselves was introduced by Fock in \cite{Fo}.
The coordinate $z_e$ associated to an edge $e$ of a triangulation
can be expressed in terms of the Penner-coordinates via
$z_e=l_a+l_c-l_b-l_d$, where $a$, $b$, $c$ and $d$ label the
other edges of the triangles that have $e$ in its boundary as indicated in the
left hand side of Figure \ref{fliplab}. 
Instead of triangulations of the Riemann surfaces
it is often convenient to consider the corresponding {\it fat graphs}, which
are defined by putting a trivalent vertex into each triangle and by 
connecting these vertices such that the edges of the triangulation are 
in one-to-one correspondence to the edges of the fat-graph.

The Teichm\"uller spaces carry a natural symplectic form, called 
Weil-Petersson symplectic form. We are therefore dealing with a 
family of phase-spaces, one for each topological type of
the Riemann surfaces. One of the crucial virtues of the 
Penner/Fock-coordinates is the fact that the Weil-Petersson symplectic
form has a particularly simple expression in these coordinates. 
The corresponding Poisson-brackets are in fact {\it constant} for
Fock's variables $z_e$,
\begin{equation}\label{poisson}
\{z_{e}^{},z_{e'}^{}\}=n_{e,e'}, \quad{\rm where}\quad
n_{e,e'}\in\{0,1,2\}.
\end{equation}

Changing the triangulation amounts to a change of coordinates for 
the (decorated) Teichm\"uller spaces. Any two triangulations can be related
to each other by a sequence of elementary moves:\\

\noindent The change of variables corresponding to the elementary move
of Figure \ref{fliplab} is easy to describe: 
\begin{equation}\label{flipfovarclass}\begin{aligned}
z_a'=& z_a-\phi(-z_e),\\
z_d'=& z_d+\phi(+z_e),
\end{aligned} 
\quad z_e'=-z_e,\quad\begin{aligned}
z_b'=& z_b+\phi(+z_e),\\
z_c'=& z_c-\phi(-z_e),
\end{aligned}\quad{\rm where}\;\;\phi(x)=\ln(e^x+1),
\end{equation}
and all other variables are left unchanged.
These transformations generate a groupoid, the Ptolemy groupoid
\cite{Pe}, that
may be abstractly characterized by generators and relations:
One has a generator $\omega_{ij}$ 
whenever the triangles labelled
by $i$ and $j$ have an edge in common. 

The mapping class group ${\rm MC}_{\Sigma}$ consists of
diffeomorphisms of the Riemann surface $\Sigma$ which are
not isotopic to the identity. Elements ${\rm MC}_{\Sigma}$ will map any
graph drawn on the surface $\Sigma$, in particular
any triangulation of $\Sigma$, into another one. Since any
two triangulations can be connected by a sequence of elementary moves
one may represent any element ${\rm MC}_{\Sigma}$ by the 
corresponding sequence of flips. It is extremely useful to think of
the algebraically complicated mapping class group ${\rm MC}_{\Sigma}$ as being
embedded into the Ptolemy groupoid.

Regarding the Teichm\"uller spaces as a collection
of phase-spaces naturally leads one to look for suitable 
Hamiltonians. A natural choice is associated to pants decompositions
of the Riemann surfaces. In the case of a Riemann surface $\Sigma_g^s$
of genus $g$
with $s$ boundary components this will be a 
collection of $3g-3+s$ closed geodesics $c_i$ on
the surface such that cutting $\Sigma_g^s$ along $c_i$, $i=1,\dots,3g-3+s$
decomposes it into a collection of three-holed spheres. The collection of  
lengths $l_i$ of the geodesics $c_i$ furnishes a set of functions
on the Teichm\"uller space $T(\Sigma_g^s)$ that Poisson-commute 
with each other, $\{l_i,l_j\}=0$ \cite{Wo}. It is possible, 
but nontrivial to represent the lengths $l_i$ as polynomials in 
the variables $e^{\pm z_e/2}$ \cite{Fo}. It seems natural to take 
$\{l_1,\dots,l_{3g-3+s}\}$ as the set of Hamiltonians to turn 
$T(\Sigma_g^s)$ into an integrable system. 

\begin{figure}
\begin{equation*}\begin{aligned}
{}& \hspace{2.5cm} {\rm A-Move}  \hspace{4.8cm} {\rm B-Move} \\
&\epsfxsize6cm\epsfbox{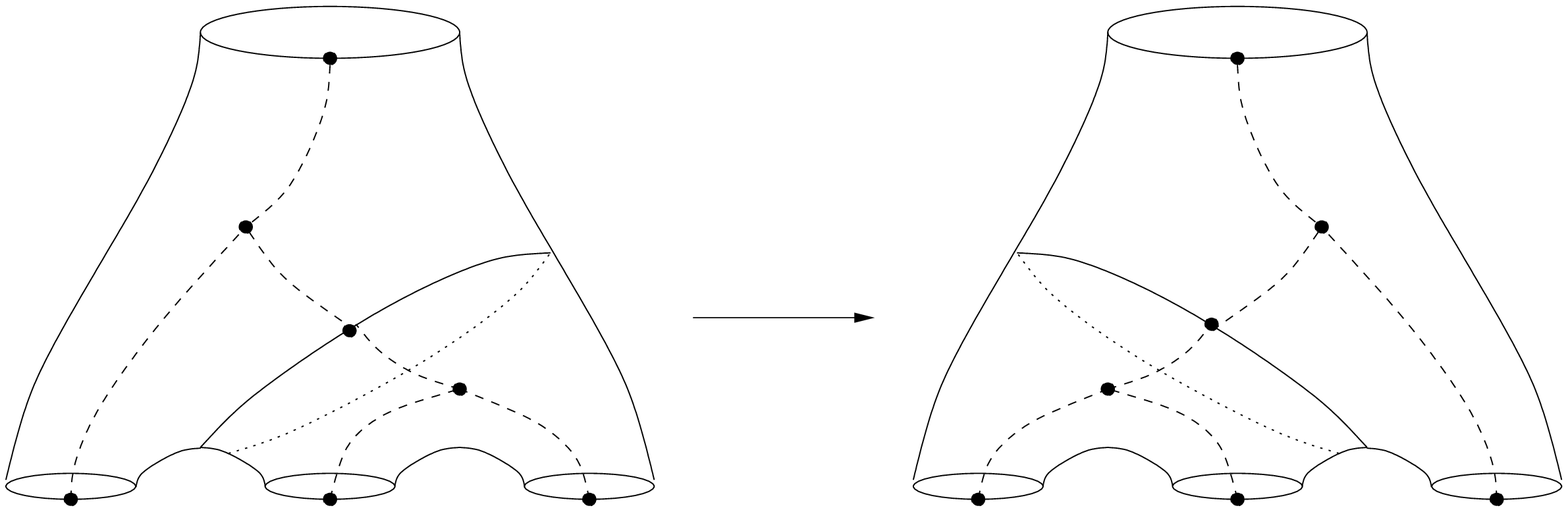} \qquad
\epsfxsize5.5cm\epsfbox{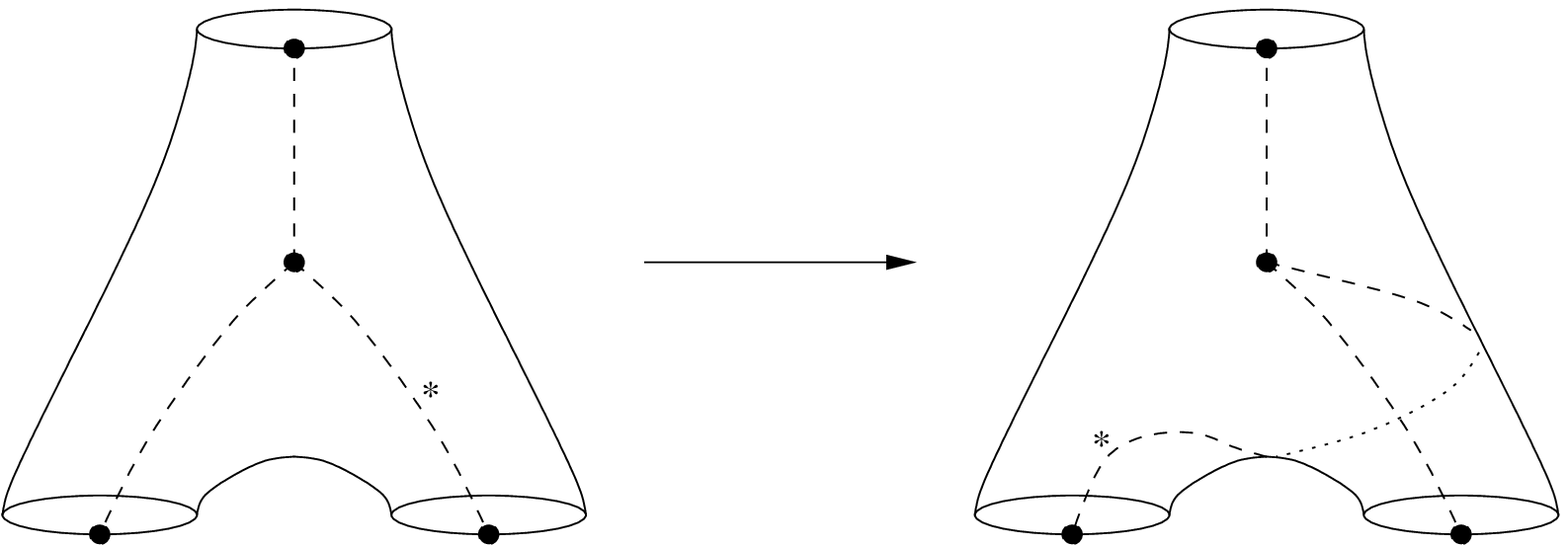}\\[1ex]
{} & \hspace{5.5cm} {\rm S-Move} \\
& \hspace{3cm}\epsfxsize6.5cm\epsfbox{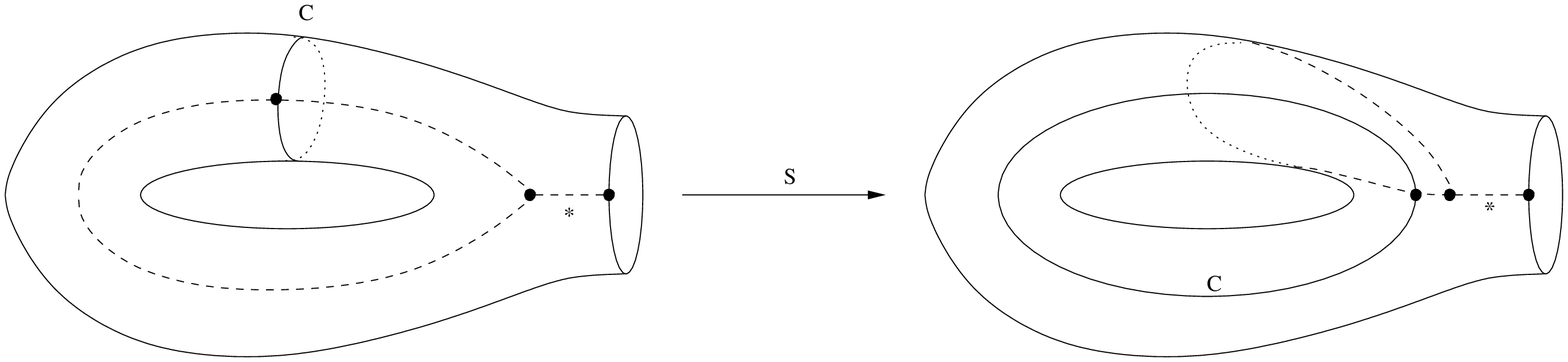}
\end{aligned}\end{equation*}
\caption{The generators of the Moore-Seiberg groupoid}\label{MS}
\end{figure}

Having chosen the $\{l_1,\dots,l_{3g-3+s}\}$ as the set of 
``action''-variables, it is amusing to note that the corresponding 
``angle''-variables are nothing but the twist-{\it angles} corresponding 
to the deformation of cutting $\Sigma_g^s$ along $c_i$ and 
twisting by some angle $\varphi_i$ before gluing back along $c_i$ \cite{Wo}.
The set
$\{l_1,\dots,l_{3g-3+s};\varphi_1,\ldots\varphi_{3g-3+s},\}$
gives another set of coordinates on the Teichm\"uller space $T(\Sigma_g^s)$,
the classical Fenchel-Nielsen coordinates. These coordinates are associated
to trivalent graphs on Riemann surfaces which are such 
that the edges that do not end in some boundary component of 
$\Sigma_g^s$ are cut by exactly one of the geodesics $c_i$.
The same type of graphs is used in the Moore-Seiberg formalism for 
conformal field theories \cite{MS1} to label spaces of conformal blocks. 
We will therefore call them Moore-Seiberg graphs. 
Again it is natural to consider the groupoid of coordinate
changes generated by the transition from one Moore-Seiberg graph
to another. A set of generators is pictorially represented 
in Figure \ref{MS}. 

This set of elementary moves generates a groupoid that will be called the 
Moore-Seiberg-groupoid. The set of relations for the 
Moore-Seiberg-groupoid is more complicated than the one 
for the Ptolemy-groupoid \cite{MS1,BK}. 

\section{Quantization of Teichm\"uller spaces}

\subsection{Algebra of observables and Hilbert space}
The simplicity of the Poisson brackets (\ref{poisson}) makes part of the
quantization quite simple. To each edge of a triangulation 
of a Riemann surface $\Sigma_g^s$ associate a quantum operator 
$\sz_e$. The algebra of observables $\CA(\Sigma_g^s)$ 
will be the algebra with generators $\sz_e$, relations 
\begin{equation}\label{comm}
[\sz_{e}^{},\sz_{e'}^{}]=2\pi i b^2 \{z_{e}^{},z_{e'}^{}\}, 
\end{equation}
and hermiticity assignment $\sz_e^{\dagger}=\sz_e^{}$. 
The algebra $\CA(\Sigma_g^s)$ 
has a center with generators
$\sfc_k$, $k=1,\ldots, s$ defined by $\sfc_k=\sum_{e\in E_k}\sz_e$
where $E_k$ is the set of edges in the triangulation that emanates from the 
$k^{\rm th}$ boundary component. 
Geometrically one may interpret $l_k$ as the geodesic length of the 
$k^{\rm th}$ boundary component \cite{Fo}\footnote{This is not 
completely obvious, though: 
The Penner coordinates are defined only for punctures, corresponding to 
$l_k\equiv 0$. However, Fock's construction \cite{Fo} of the  
variables $z_e$ also works for surfaces with geodesic boundary, in which case
the $l_k$ indeed measure the lengths of its components.}. 
The representations of $\CA(\Sigma_g^s)$ that we are going to consider
will therefore be such that the generators $\sfc_k$ are 
represented as the operators of multiplication by real 
positive numbers $l_k$.
The tuple $\Lambda=(l_1,\dots,l_s)$ of lengths of the boundary components
will figure as a label of the representation 
$\pi(\Sigma_g^s,\Lambda)$ of the algebra $\CA(\Sigma_g^s)$.
   
To complete the definition of the 
representation $\pi(\Sigma_g^s,\Lambda)$ by operators 
on a Hilbert space $\CH(\Sigma_g^s)$ one just needs to find 
linear combinations $\sq_1,\dots,\sq_{3g-3+s}$ and 
$\spp_1,\dots,\spp_{3g-3+s}$ of the 
$\sz_{e}$ that satisfy $[\sq_i,\spp_j]=2\pi i b^2\delta_{ij}$.
The representation of $\CA(\Sigma_g^s,\Lambda)$ on 
$\CH(\Sigma_g^s):=L^2(R^{3g-3+s})$ is then 
defined by choosing 
the usual Schr\"odinger representation for the $\sq_i$, $\spp_i$.  

\begin{figure}
\epsfxsize7cm
\centerline{\epsfbox{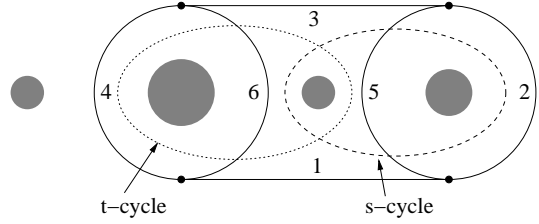}}
\caption{Fat graph for the sphere with four holes (shaded) 
with numbered edges.}
\label{fourpt}\end{figure}

Let us discuss the example of a sphere with four holes. 
We shall consider the fat graph drawn in Figure \ref{fourpt} above.
The algebra $\CA(\Sigma_{0}^{4})$
has six generators $\sz_i$ $i=1,\dots 6$ with nontrivial relations
$[\sz_i,\sz_j]=2\pi i b^2$ for 
\begin{equation}
\begin{aligned}(i,j)\;\in\;\{ & 
(1,2)\; ,\; (1,6) \; ,\; (2,3)\; ,\; (2,5)\; ,\;(3,4)\; ,\;(3,5)\\
& (4,1)\; ,\; (4,6) \; ,\; (5,2)\; ,\; (5,1)\; ,\;(6,3)\; ,\;(6,4)\;\}.
\end{aligned}
\end{equation}
The four central elements corresponding to the holes in $\Sigma_{0,4}$ are 
\begin{equation}
\begin{aligned}
\sfc_1\;=\;& \sz_4+\sz_6,\\
\sfc_3\;=\;& \sz_2+\sz_5,
\end{aligned}\qquad\begin{aligned}
\sfc_2\;=\;& \sz_1+\sz_3+\sz_5+\sz_6,\\
\sfc_4\;=\;& \sz_1+\sz_2+\sz_3+\sz_4.
\end{aligned}
\end{equation}
After fixing the lengths of the four holes one is left with two variables,
say $\sq\equiv \sz_3$ and $\spp\equiv \sz_5$. Choosing the Schr\"odinger
representation for $\sq$, $\spp$ one simply finds $\CH(\Sigma_0^4)
\simeq L^2(\BR)$. 

\subsection{Representation of the Ptolemy groupoid}

The first task is to find the quantum counterpart of the 
change of variables corresponding to the elements
of the Ptolemy groupoid. A handy formulation of 
the solution \cite{Fo,CF} can be given in terms 
of the Fock-variables: The change of variables
corresponding to the
elementary move depicted in Figure \ref{fliplab} is given by
\begin{equation}\label{flipfovar}\begin{aligned}
\sz_a'=& \sz_a-\phi_{b}(-\sz_e),\\
\sz_d'=& \sz_d+\phi_{b}(+\sz_e),
\end{aligned} 
\qquad \sz_e'=-\sz_e,\qquad\begin{aligned}
\sz_b'=& \sz_b+\phi_{b}(+\sz_e),\\
\sz_c'=& \sz_c-\phi_{b}(-\sz_e),
\end{aligned}
\end{equation}
where the special function $\phi_{b}(x)$ 
is defined as
\begin{equation}
\phi_b(z)\;=\;\frac{\pi b^2}{2}
\int\limits_{i0-\infty}^{i0+\infty}dw
\frac{e^{-i zw}}{\sinh(\pi w)
\sinh(\pi b^{2}w)}.
\end{equation}
$\phi_{b}(x)$ represents the quantum deformation of the
classical expression given in (\ref{flipfovarclass}).
The formulae (\ref{flipfovar}) define a representation of the Ptolemy groupoid
by automorphisms of $\CA(\Sigma_g^s)$, see \cite{CF}.

Let us recall that the mapping class group ${\rm MC}_g^s$ can be embedded
into the Ptolemy groupoid. Having realized the latter therefore 
gives a representation of ${\rm MC}_g^s$ by automorphisms of 
$\CA(\Sigma_g^s)$.
However, it turns out that one has to deal with 
subtleties related to projective phases if one wants to find a 
consistent representation of the
mapping class by operators on $\CH(\Sigma_g^s)$.
An elegant solution to this problem
was given by Kashaev in \cite{Ka1}. It
uses an enlarged set of variables, where
pairs of variables are associated to the 
$2M=4g-4+2s$ triangles of a triangulation instead
of its edges. The reduction to $(\CA(\Sigma_g^s),\CH(\Sigma_g^s))$ 
can be described
with the help of a simple set of constraints \cite{Ka1}. 
Kashaev's formalism
is particularly useful to control the projective phases in the 
relations of the mapping class group \cite{Ka2}.
It turns out that
the mapping class group is realized only projectively \cite{Ka2}.

\subsection{The length operators}  

Our aim is to make contact with the Liouville conformal field theory.
In the Moore-Seiberg formalism for conformal field theories one 
considers bases for the space of conformal blocks that are associated
to pants decompositions of Riemann surfaces. Our task may be seen
as a quantum counterpart of the task to construct the
change of variables from the Penner- to the Fenchel-Nielsen
coordinates.

\begin{figure}[b]
\epsfxsize3cm
\centerline{\epsfbox{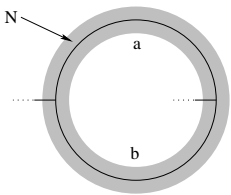}}
\caption{Neighborhood $N$ and the part of the fat graph $\Gamma$
contained in $N$}
\label{cycle}\end{figure}

The first problem to address is of course the construction of 
quantum counterparts for the geodesic length functions on Teichm\"uller
space. For a subset of the closed geodesics $\gamma$ on 
a Riemann surface $\Sigma$ one may find a fat graph $\Gamma$ 
that in an annular
neighborhood $N$ of $\gamma$ looks as drawn in Figure \ref{cycle}. 
The subset of closed geodesics for which this is possible includes
all closed geodesics in a Riemann surface of genus zero and all
non-separating cycles \cite{TT}.
In this case one finds according to \cite{Fo} an expression for the
hyperbolic cosine of the geodesic length function that is easy to quantize,
\begin{equation}\label{lodef}
\SL_{\gamma}\;=\;2\cosh2\pi b\spp+e^{2\pi b \sx},\qquad
\begin{aligned}
2\pi b \sx \;=&\;\fr{1}{2}(\sz_a-\sz_b), \\
2\pi b \spp\;=&\;\fr{1}{2}(\sz_a+\sz_b).
\end{aligned}
\end{equation}
This finite difference operator can be shown to be self-adjoint 
and to have a nondegenerate spectrum given by the interval $(2,\infty)$
\cite{Ka3}. It follows that there exists a self-adjoint
operator $\sll_{\ga}$ with spectrum $\BR_+$ 
such that $\SL_{\ga}=2\cosh\frac{1}{2}\sll_{\ga}$. 
$\sll_{\ga}$ is the quantum operator corresponding to the 
hyperbolic length around the geodesic $\ga$.

The representation of $\SL_{\ga}$ in terms of the Fock variables
associated to another fat graph $\Gamma'$ can become complicated. 
It has to be found by applying the automorphism of $\CA(\Sigma_g^s)$
that corresponds to the element of the Ptolemy groupoid which relates
$\Gamma$ and $\Gamma'$. As an example let us quote the expressions for
the operators $\SL_s$ and $\SL_t$ 
representing the lengths of the geodesics isotopic 
to the s- and t-cycles drawn in Figure \ref{fourpt} respectively. 
\begin{equation}\begin{aligned}
\SL_s\;=\;& 2\cosh\big(\spp+\fr{1}{2}(\sfc_4-\sfc_3)\big)\\
& +e^{-\frac{1}{2}\sq}
\Big[ 2\cosh\fr{1}{2}\big(\spp+\fr{1}{2}(\sfc_1+\sfc_4-\sfc_2-\sfc_3)\big)
2\cosh\fr{1}{2}\spp \Big]e^{-\frac{1}{2}\sq}\\
\SL_t\;=\;& 2\cosh\big(\spp-\fr{1}{2}(\sfc_3-\sfc_2)\big)\\
& +e^{+\frac{1}{2}\sq}
\Big[2\cosh\fr{1}{2}\big(\spp-\fr{1}{2}(\sfc_1+\sfc_2+\sfc_3-\sfc_4)\big)
2\cosh\fr{1}{2}(\spp-\sfc_3)\Big]e^{+\frac{1}{2}\sq}.
\end{aligned}
\end{equation}
Complete sets of eigenfunctions for the operators $\SL_s$ and $\SL_t$
were found in \cite{PT}. They will be denoted by
\begin{equation}\label{efnotation}
\Psi^s_{l}\bigl[\begin{smallmatrix}l_3 & l_2\\l_4 & l_3\end{smallmatrix}
\bigr](q) \quad\text{and}\quad
\Psi^t_{l}\bigl[\begin{smallmatrix}l_3 & l_2\\l_4 & l_3\end{smallmatrix}
\bigr](q)\quad\text{respectively}.
\end{equation}

It will be shown \cite{TT} that the definition of the length operators 
$\sll_{\ga}$ can be extended to {\it arbitrary} closed geodesics $\ga$
such that \begin{itemize}
\item $\sll_{\ga}$ is self-adjoint with spectrum $\BR_+$,
\item $[\sll_{\ga},\sll_{\ga'}]=0$ if $\ga\cap\ga'=\emptyset$.
\end{itemize}
Moreover, diagonalization of the length operator  
$\sll_{\ga}$ for a closed geodesic $\ga\subset\Sigma$ leads to a 
{\it factorization} of $\pi(\Sigma,\Lambda)$ 
in the following sense. Let $\Sigma'_{\ga}\equiv
\Sigma\setminus\ga$ be the possibly
disconnected Riemann surface obtained by cutting along $\ga$.
The coloring $\Lambda$ of the boundary components of $\Sigma$ can be
naturally extended to a coloring 
$\Lambda'_{\ga,l}$ for $\Sigma'_{\ga}$ by assigning the number $l\in\BR^+$ to
the two new boundary components that were created by
cutting along $\ga$. The spectral representation for $\sll_\ga$ then
yields the following representation for $\pi(\Sigma,\Lambda)$.
\begin{equation}
\pi(\Sigma,\Lambda)\;\simeq\; \int_{\BR^+}^{\oplus}dl \;
\pi(\Sigma'_{\ga},\Lambda'_{\ga,l}).
\end{equation}
The corresponding representations of the mapping class group 
factorize/restrict accordingly \cite{TT}.
This allows one to construct bases (in the sense of generalized functions)
for $\CH(\Sigma)$ labelled by the assignments of lengths to the 
closed geodesics $c_1,\dots,c_{3g-3+s}$ that define a pants 
decomposition. 

\subsection{Realization of the Moore-Seiberg groupoid}

Thanks to the factorization properties of the quantized Teichm\"uller
spaces one may indeed construct 
a realization of the Moore-Seiberg groupoid
by associating unitary operators to the elementary moves depicted in
Figure \ref{MS} \cite{TT}. 

In the case of the A-move, the unitary operator will simply be the
one that describes the change of basis between 
the eigenfunctions of the length operators for s- and
t-cycles of the four-holed sphere (see Figure \ref{fourpt}) respectively.
The unitary operator representing the A-move has
matrix elements $F_{ll'}$ given by the overlap of $\Psi^s_l$ and
$\Psi^t_{l'}$,
\begin{equation}
F_{ll'}^{\rm T}
\bigl[\begin{smallmatrix}l_3 & l_2\\l_4 & l_3\end{smallmatrix}\bigr]
\;=\;
\int\limits_{\BR}dq\;\,
\overline{
\Psi^t_{l}\bigl[\begin{smallmatrix}l_3 & l_2\\l_4 & l_3\end{smallmatrix}
\bigr](q)}\;
\Psi^s_{l}\bigl[\begin{smallmatrix}l_3 & l_2\\l_4 & l_3\end{smallmatrix}
\bigr](q).
\end{equation}
The matrix elements $F_{ll'}$ essentially 
coincide with the b-Racah-Wigner coefficients 
for the quantum group $\CU_q(\fsl(2,\BR))$ studied and calculated 
in \cite{PT}.

A realization of the B-move in terms of the generators of the 
Ptolemy groupoid was already discussed in \cite{Ka3}, where 
it was found to be related to the R-operator for $\CU_q(\fsl(2,\BR))$
that was proposed by Faddeev in \cite{Fa}. It turns out \cite{TT}
that the corresponding operator becomes diagonal in a basis
that diagonalizes the three
length operators $\sll_i$, $i=1,2,3$ corresponding to the boundaries 
of the three-holed sphere with eigenvalues $l_i$. 
A quick way to find the expression
for the eigenvalue $B(l_3,l_2,l_1)$
of the operator that represents the B-move
is to combine the observation from \cite{Ka3} with \cite[Theorem 6]{BT}.
The result is
\begin{equation}
B^{\rm T}(l_3,l_2,l_1)\;=\;
\exp\bigl(\pi i(\Delta_{l_3}-\Delta_{l_2}-\Delta_{l_1})\bigr),
\qquad \Delta_l=\frac{1}{4b^2}\Bigl((1+b^2)^2
+\Bigl(\frac{l}{2\pi}\Bigr)^2\Bigr).
\end{equation}

The data $(F^{\rm T},B^{\rm T})$ turn out \cite{TT} to be sufficient
to construct a projective representation
of the Moore-Seiberg groupoid on the quantized Teichm\"uller
spaces $\CH_g^s$. The expression for the generator 
of the S-move in terms of $(F^{\rm T},B^{\rm T})$ is similar to the
one that was found for rational conformal field theories in \cite{MS2}.

\section{Relation to Liouville theory}

The spectrum of quantum Liouville theory can be represented as follows:
\begin{equation}\label{spec}
\CH^{\rm L}\;\simeq\; \int_{\BS}d\alpha \;\CV_{\alpha}\otimes\bar{\CV}_{\alpha}
,\qquad\BS\equiv \frac{Q}{2}+i\BR^+.
\end{equation}
In (\ref{spec})
we used the notation $\CV_{\alpha}$, $\bar{\CV}_{\alpha}$ for the  
unitary highest weight representations   
of the Virasoro algebras 
which are generated from the modes of the holomorphic and 
anti-holomorphic parts
of the energy-momentum tensor respectively. 
The central charge of the representations is given by the parameter
$b$ via $c=1+6Q^2$, $Q=b+b^{-1}$, and the
highest weight of the representations $\CV_{\alpha}$, $\bar{\CV}_{\alpha}$ is 
parametrized as $\Delta_{\alpha}=\alpha(Q-\alpha)$.

Quantum Liouville theory in genus zero may be characterized by the 
set of n-point functions of the primary fields $V_{\alpha}(z,\bar{z})$
which are the quantized exponential functions $e^{2\alpha\phi(z,\bar{z})}$
of the Liouville field.
The construction \cite{TL} of n-point functions 
$\langle V_{\alpha_n}(z_n,\bar{z}_n)\dots V_{\alpha_1}(z_1,\bar{z}_1)
\rangle$  can be conveniently formulated within the
Moore-Seiberg formalism. A basis for the 
space of Liouville 
conformal blocks can be associated to each pants decomposition of the 
n-punctured sphere. 
The elements of this basis are 
constructed by taking matrix elements of compositions of chiral
vertex operators \cite{TL}. A labelling of the 
elements of such a basis will be obtained
by assigning numbers $\alpha_i\in\frac{Q}{2}+i\BR^+$ to the curves $c_i$ that
define the pants decomposition. 
Changing the pants decomposition simply amounts to a change of basis
in the space of conformal blocks. The corresponding transformations 
can again be constructed out of the representatives for 
the elementary A-moves and B-moves depicted in Figure \ref{MS}. 
These moves are represented in terms of the
fusion coefficients 
\newcommand{\al}{\alpha}
$F_{\al\al'}^{\rm L}
\bigl[\begin{smallmatrix}\al_3 & \al_2\\ \al_4 & \al_3\end{smallmatrix}\bigr]$
and the elementary braiding phase 
$B^{\rm L}(\al_3,\al_2,\al_1)$. 
The explicit expressions for the data $(F^{\rm L},B^{\rm L})$ 
were determined in \cite{TL}. 
We would like to 
emphasize that the definition of the data $(F^{\rm L},B^{\rm L})$ 
involves nothing but the representation theory of the Virasoro algebra.

The main observation to be made is the following. Upon choosing 
a natural normalization for the Liouville conformal blocks
one finds 
\begin{equation}
(F^{\rm T},B^{\rm T})\;\equiv\; (F^{\rm L},B^{\rm L})
\quad \text{provided that}\quad \al_i=\frac{Q}{2}+i\frac{l_i}{4\pi b}.
\end{equation}
It follows that the spaces of conformal blocks of Liouville theory  
and the Hilbert spaces from the quantization of the Teichm\"uller spaces
are indeed isomorphic as representations of the mapping class group.
So far the conformal blocks of Liouville theory were only constructed
in genus zero, but our results also imply that the 
corresponding mapping class group representation has a 
consistent and essentially unique extension to higher genus. Moreover,
the geodesic length $l$ of a closed geodesic on the Riemann surface $\Sigma$
is indeed directly related to the representation $\CV_{\alpha}$ 
that ``flows through that cycle'' as was anticipated in \cite{V}.


\end{document}